# Stimulation of human red blood cells leads to $Ca^{2+}$-mediated intercellular adhesion

Short title: A novel RBC adhesion mechanism

Patrick Steffen[1], Achim Jung[1], Duc Bach Nguyen[2], Torsten Müller[3], Ingolf Bernhardt[2],

Lars Kaestner[4],* and Christian Wagner[1],*

1   Experimental Physics Department, Saarland University, 66123 Saarbruecken, Germany

2   Biophysics, Saarland University, 66123 Saarbruecken, Germany

3   JPK Instruments AG, Bouchéstrasse 12, 12435 Berlin, Germany

4   Molecular Cell Biology, Saarland University, 66421 Homburg, Germany

\* Corresponding authors:

| | |
|---|---|
| Prof. Christian Wagner | Dr. Lars Kaestner |
| Experimental Physics Department | Institute for Molecular Cell Biology |
| Building E 2 6 | Building 61 |
| 66123 Saarbruecken | 66421 Homburg/Saar |
| e-mail: c.wagner@mx.uni-saarland.de | anlkae@uks.eu |
| phone: +49 681 302 3003 | +49 6841 1626 103 |
| fax:     +49 681 302 4676 | +49 6841 1626 104 |

word count: 6691

Abstract

Red blood cells (RBCs) are a major component of blood clots, which form physiologically as a response to injury or pathologically in thrombosis. The active participation of RBCs in thrombus solidification has been previously proposed but not yet experimentally proven. Holographic optical tweezers and single-cell force spectroscopy were used to study potential cell-cell adhesion between RBCs. Irreversible intercellular adhesion of RBCs could be induced by stimulation with lysophosphatidic acid (LPA), a compound known to be released by activated platelets. We identified $Ca^{2+}$ as an essential player in the signaling cascade by directly inducing $Ca^{2+}$ influx using A23187. Elevation of the internal $Ca^{2+}$ concentration leads to an intercellular adhesion of RBCs similar to that induced by LPA stimulation. Using single-cell force spectroscopy, the adhesion of the RBCs was identified to be approximately 100 pN, a value large enough to be of significance inside a blood clot or in pathological situations like the vasco-occlusive crisis in sickle cell disease patients.

# Introduction

Previous studies of the interactions between red blood cells (RBCs) have relied either on hydrodynamic interactions [1] or interactions mediated by plasma macromolecules. Co-adhesion of RBCs under physiological conditions is known as rouleaux formation. These structures might be generated either by plasma polymers bridging between the cells [2, 3] or, more likely, by depletion forces. The adhesion energies are small enough to allow for breakage in the shear flow, and the effect is thought to be reversible, even if some controversial data exist [4]. This reversibility is crucial for the ability of RBCs to cross capillary sections that are smaller than the cell diameter. However, in several pathologies, such as sickle cell disease or thalassemia, an increased propensity for red blood cell-cell adhesion has been observed [5]. Clot formation is a collective phenomenon based on the interplay of many components. In current models, the contribution of RBCs to the clotting process is thought to be purely passive, i.e., they are simply caged in the fibrin network due to their prevalence in the blood. Based on experimental and clinical studies that have shown a correlation between decreased hematocrit values and longer bleeding times [6-8] and their own experiments, Andrews and Low summarized the evidence for active participation of RBCs in thrombus formation. Kaestner et al. [9] proposed a signaling mechanism based on $Ca^{2+}$ entry via a nonselective voltage-dependent cation (NSVDC) channel that is permeable to mono- and bivalent cations like $Na^+$, $K^+$ and $Ca^{2+}$ [10, 11]. This channel has been shown to be activated by prostaglandin $E_2$ ($PGE_2$), lysophosphatidic acid (LPA) [9, 12] and by the mechanical deformations that occur when RBCs pass through capillaries smaller than their resting diameter [13]. $PGE_2$ and LPA are extracellular, local mediators that are released both by platelets after their activation within the coagulation cascade and by RBCs themselves under mechanical $Ca^{2+}$ stress [14]. The aim of this study was to test the hypothesis that intercellular RBC adhesion can occur directly (without participation of platelets). To perform this at the level of individual cells, we used the complementary methods of non-invasive holographic optical tweezers (HOT) [15], using the momentum of light and single-cell force spectroscopy to quantify the occurring adhesion.

# Materials and methods

## RBC preparation and fluorescence microscopy

For experiments using the optical tweezers and microfluidics, fresh blood from healthy donors was obtained by a fingertip needle prick, while for annexinV-FITC fluorescence measurements, human venous blood was drawn from healthy donors. Heparin or EDTA was used as an anticoagulant. The blood was obtained within one day of the experiment from the Institute of Clinical Haematology and Transfusion Medicine of the Saarland University Hospital. The cells were washed three times by centrifugation (2000 g, 3 min) in a HEPES buffered solution of physiological ionic strength (PIS-solution) containing the following (in mM): 145 NaCl, 7.5 KCl, 10 glucose and 10 HEPES, pH 7.4, at room temperature. The buffy coat and plasma were removed by aspiration. For $Ca^{2+}$ imaging, RBCs were loaded with 5 µM Fluo-4 AM (Molecular Probes, Eugene, USA) from a 1 mM stock solution in Dimethyl sulfoxide (DMSO) with 20% Pluronic (F-127, Molecular Probes). Loading was performed in 1 mL of PIS-solution at an RBC haematocrit of approximately 1% for 1 h at 37 °C. The cells were washed by centrifugation once more and equilibrated for de-esterification for 15 min. LPA, prepared from a stock solution of 1 mM in distilled water, and the $Ca^{2+}$ ionophore A23187, prepared from a stock solution of 1 mM in ethanol, were obtained from Sigma Aldrich (St. Louis, USA). To investigate phosphatidylserine (PS) exposure, cells were stained with annexinV-FITC (Molecular Probes). Annexin V-FITC was delivered in a unit size of 500 µL containing 25 mM HEPES, 140 mM NaCl, 1 mM EDTA, pH 7.4, and 0.1% bovine serum albumin. Five hundred microliters of annexin binding buffer (10 mM HEPES, 140 mM NaCl, 2.5 mM $CaCl_2$, pH 7.4) was added to 1 µL of washed, packed RBCs previously treated with LPA or A23187. Afterwards, 5 µL annexin V-FITC was added, and the cells were mixed gently. The probes were incubated at room temperature in the dark for 15 min and then washed 2 times in 1 mL of annexin binding buffer (2000 g, 3 min). Finally, the cells were re-suspended in 500 µL of annexin binding buffer and used for microscopy. The measurements were obtained from images taken with a CCD camera (CCD97, Photometrics, USA).

## Holographic optical tweezers (HOT)

A schematic drawing of our experimental setup is shown in supplemental Fig. S1a. A Nd:YAG infrared laser (Ventus, Laser Quantum, Stockport, UK) with a beam width of 2.5 mm was coupled for coarse alignment with a visible He-Ne laser via a dichroic mirror. The latter was switched off during the experiments. Both beams were expanded five-fold (BM.X, Linos Photonics, Göttingen, Germany) to overfill the 8 mm back aperture of the microscope objective (CFI Plan Fluor 60x oil immersion, Nikon Corp., Tokyo, Japan). The optics were integrated into an inverted fluorescence microscope (TE-2000, Nikon Corp.). This allowed for combined trapping and fluorescence or differential interference contrast (DIC) measurements. Images were taken with an electron multiplication CCD camera (Cascade 512F, Roper Scientific, Trenton, USA) with a typical frame rate of 100 Hz. To reduce vibrations, the original cooling fan was replaced with water-cooling components. In addition, the optical setup was placed on active vibration isolation elements (Vario Series, Halcyonics, Göttingen, Germany). The phase of the laser's electric field was modified using a spatial light modulator (PPM X8267-15, Hamamatsu Photonics, Hamamatsu City, Japan) to create the desired trap pattern in the focal plane of the microscope objective. In order to test for RBC adhesion after stimulation, a suitable configuration of independently movable optical traps was created (see the schematic sketch in Fig. 1a ). The laser power in each trap was approximately 5 mW. Cells could be moved against one another by replaying a series of kinoforms on the computer-controlled spatial light modulator. As an example, supplemental Fig. S1b shows a quadrotrap in parallel and perpendicular configuration for RBC arrangement along with the corresponding kinoforms.

## Microfluidics

In order to provide a controlled yet interchangeable solvent environment, a microfluidic polydimethylsiloxane (PDMS) cell was constructed by standard soft lithography (inset in Fig. 2). The cell consists of two inlets from which either the cells in the buffer solutions or LPA or the ionophore A23187 (final concentration of 40 µM) could be applied. The flow was driven by a hydrostatic pressure difference that allowed for fine tuning. After injection of a new sample, the flow could be brought approximately to rest by eliminating the pressure difference. Behind the y-junction, the flow remained laminar, and the two solutions did not mix, except for very slow diffusion. The RBCs could then be captured by the HOT and transferred to the new environment. Thus, the cells could be transferred from one solvent condition to another in a rapid and controlled manner. The same measurements were also conducted in a Petri dish as follows: RBCs were incubated for 10 min with either A23187 or LPA in the presence of $Ca^{2+}$ (for concentration values, please refer to the figure legends), followed by a fast sequence of adhesion tests between various cells. There were more than 50 cells in each experiment; the total number of tested cells amounted to 250.

## Single-cell force spectroscopy

In order to quantify the occurring cell-cell adhesion, an atomic force microscope (AFM) (CellHesion 200 with increased pulling range up to 100 µm, JPK Instruments, Germany) was used to conduct single-cell force spectroscopy measurements. In these measurements, a single RBC was attached to an AFM cantilever with the help of an appropriate adhesive. Cell-Tak$^{TM}$ (BD Biosciences) turned out to be the most efficient adhesive for binding RBCs to the cantilever. A stock solution of 2.4 mg/mL was diluted 1:30 according to instructions from BD. The cantilevers were incubated for at least 30 min at room temperature and used after rinsing with PIS solution. After the cell capture (CellTak$^{TM}$ functionalized cantilevers brought in contact with a 0.2 nN force for 10 s), the approach- and retraction speed were set to 5 µm/s. The pulling range varied between 30 and 50 µm, and the contact time varied from 0 to 120 s. During spectroscopy experiments, the deflection of the cantilever was monitored in real time using a built-in feature of the AFM software. The spring constant of the cantilever was determined by the common thermal noise method. The cantilevers used were tipples TL1 with a nominal spring constant of 0.03 N/m (Nanoworld). The cell morphology was monitored using phase contrast microscopy. In the course of the experiment, the cantilever with the attached cell was lowered onto another cell attached to the substrate, which was coated with 0.005% Poly-L-Lysine, until a preset defined constant force was reached and kept stationary for a defined contact time. The conditions during contact were determined by the choice of the particular closed-loop mode, specifically at a constant piezo position, after reaching a prescribed maximum pushing force of 400 pN. Subsequently, the cantilever was withdrawn at a constant speed. During approach and retraction, the cantilever deflected as a consequence of the acting forces. This deflection, which is proportional to the acting forces, was recorded in force-distance curves. The retraction curve was typically characterized by the maximum force required to separate the cells from each other, referred to as the maximum unbinding force $F_{max}$.

## Statistical significance

A Student's t-test was used to test the results obtained from the adhesion experiments for statistical significance. Statistical significance of the data were defined as follows: $p > 0.05$ (n.s.), $p \leq 0.05$ (*); $p \leq 0.01$ (**), $p \leq 0.001$ (***).

# Results

## Red blood cell stimulation with LPA

As pointed out in the Introduction, RBCs can be stimulated by LPA, and this has been proposed to contribute to the active participation of RBCs in the later stage of thrombus formation. In order to test for altered intercellular adhesion behavior, we set up holographic optical tweezers (for details, refer to supplemental Fig. S1). In the blood flow, cell-cell contact times are rather short when RBCs "bump" into each other. To mimic this physiological condition, two cells were grabbed with the laser foci (compare Fig. 1c) and moved back and forth as depicted in Fig. 1a and supplemental movie SM1, in a cyclical manner. Upon stimulation with 2.5 μM LPA, the RBCs adhered to each other as plotted and visualized in an image sequence in Figs. 1b and 1c, respectively (the movie is available in the supplemental material.) During the stimulation procedure, most of the RBCs remained in their discocyte shape. The separation force could not be determined by the HOT approach because it exceeds the force of the laser tweezers. In this configuration, 72 % of the cells tested showed such an irreversible adhesion (grey bars in Fig. 2b). To exclude any dependencies on the interaction surface due to the anisotropic shape of the cells, we aimed for another condition using spherocytes. This was realized by increasing the LPA concentration to 10 μM, which is still within the physiologically observed range. In parallel to the adhesion experiments, the $Ca^{2+}$ uptake of the cells was followed up by Fluo-4 imaging, as the trace in Fig. 2a shows for the stimulation with LPA. Additionally, a microfluidic system, schematically plotted as an inset in Fig. 2a, allowed a fast change of media by moving the RBCs using the holographic optical tweezers from one "channel" to the other. In this way, we tested cells under five different conditions: (i) PIS-solution, (ii) PIS-solution containing 2 mM $Ca^{2+}$ and 10 μM LPA, (iii) PIS-solution containing 2 mM $Ca^{2+}$ and 2.5 μM LPA, (iv) PIS-solution containing 2 mM $Ca^{2+}$ and no LPA, and finally (v) PIS-solution containing 2 mM EDTA and 10 μM LPA. We used at least 60 cells per condition. The results are summarized in Fig. 2b. The RBC stimulation with LPA (2.5 μM as well as 10 μM) in the presence of extracellular $Ca^{2+}$ led to an immediate qualitative change in the adhesion behavior: cells irreversibly stuck to each other. Additionally, RBCs were stained with an annexinV-FITC conjugate to probe for phosphatidylserine (PS) exposure of the RBCs by means of fluorescence microscopy. In Fig. 2c, we show that the 2.5 μM LPA-induced $Ca^{2+}$ influx indeed was associated with PS transport from the inner leaflet to the outer leaflet of the RBC membrane. This staining was not observed in cells treated with conditions (i), (iv) and (v).

## Approaching the signaling entities

The initial stimulation experiments using LPA revealed that an LPA-induced $Ca^{2+}$ influx leads to intercellular RBC adhesion. To test whether this is a pure $Ca^{2+}$ effect or if the presence of LPA is required, experiments were performed using the $Ca^{2+}$ ionophore A23187 as an artificial tool to increase the intracellular $Ca^{2+}$ concentration. After transferring the Fluo-4 loaded RBCs to the A23187 solution, the Fluo-4 fluorescence signal increased almost immediately, i.e., faster than after application with LPA. The $Ca^{2+}$ influx into the cell is depicted in Fig. 3a. As illustrated in Fig. 3b, during the $Ca^{2+}$ increase, the cells undergo a shape transformation from discocytes to spherocytes via an intermediate step of echinocytes. Testing for adhesion was performed using a procedure identical to the LPA experiments performed in the following four media: (i) PIS-solution, (ii) PIS-solution containing 2 mM $Ca^{2+}$ and 40 µM A23187, (iii) PIS-solution containing 2 mM $Ca^{2+}$ and no A23187, and (iv) PIS-solution containing 2 mM EDTA and 40 µM A23187. The results are summarized in Fig. 3c. Intercellular adhesion was significantly increased compared to the controls only when the intracellular $Ca^{2+}$ concentration was increased. There have been controversial reports about PS transition to the outer membrane leaflet that claim merely coincidental increase in $Ca^{2+}$ that is not related to the PS exposure [16, 17]. Therefore, RBCs were stained with annexin-V-FITC conjugates in order to visualize PS exposure on the surface of RBCs. As depicted in Fig. 3d, treatment with A23187 in the presence of extracellular $Ca^{2+}$ leads to a clear annexin-V binding (PS staining) on the outer leaflet of the membranes. Under all the other conditions (i), (iii), and (iv), no such staining could be observed.

## Quantification of the intracellular adhesion

To allow a discussion of a physiological (or pathophysiological) relevance of the described adhesion process, one needs to determine the separation force. As described above, the separation force exceeds the abilities of the HOT. Therefore, single-cell force spectroscopy [18] was utilized for determination of the force. Two different measurements were conducted: control measurements in which the cells remained untreated and measurements in which the cells were treated with a concentration of 2.5 µM LPA. Example curves for both experiments are depicted in Fig. 4b. It is an inherent part of the single-cell force spectroscopy procedure that cells need to be brought in contact by a certain force application (step 2 in Fig. 4a). In the control measurements, it was only possible to detect a weak interaction between the cells, whereas in the LPA measurements, a pronounced adhesion of the red blood cells could be observed. The step-wise release plotted in Fig. 4b (step 3 in Fig. 4a) was typical for all the cells measured. All results of the measurements are collected into histograms summarized in Figs. 4c and 4d. The mean value of the maximum unbinding force of untreated red blood cells amounted to $28.8 \pm 8.9$ pN (s.d.) (n=71), whereas in the LPA experiments, the mean value of the maximum unbinding force amounted to a much higher value of $100 \pm 84$ pN (s.d.) (n=193, from three different donors) indicating a severe difference in adhesion behavior of untreated and LPA-stimulated RBCs.

# Discussion

## LPA stimulation leads to intercellular adhesion

It is an established fact that the stimulation of RBCs with LPA leads to a $Ca^{2+}$ influx through an NSVDC channel [9, 12]. It is further known that the increased intracellular $Ca^{2+}$ results in the stimulation of the $Ca^{2+}$-activated $K^+$ channel (Gardos channel) [19, 20] and the activation of the lipid scramblase [21, 22] (see below for further discussion). Based on these mechanisms, an increased aggregation tendency of stimulated RBCs has been hypothesized [9, 23]. The development of the HOT provides a tool for testing this hypothesis at the cellular level. Concentrations of 2.5 μM and 10 μM LPA were chosen because they seemed to be within the common range of concentrations used with other cell types [24, 25] in addition to RBCs [17]. Moreover, this concentration is comparable to the local LPA concentration in the immediate surroundings of activated platelets, e.g., inside a blood clot [26, 27]. While the choice of the LPA concentration did not seem to have any significant effect on the adhesion rate itself, it had an impact on the shape of the RBCs. This relationship is discussed in the next section. After an intracellular $Ca^{2+}$ increase was observed by fluorescence imaging, the setup mode was switched to white light for better HOT operation. Then, the cells were brought into contact and adhered to each other immediately. Because the time for the $Ca^{2+}$ increase varied from cell to cell, which is in agreement with previous investigations [9, 28], the time between the initial stimulation of the cell and the final adhesion varied between 10 and 140 s, yielding an average value of 72.75 ± 46.79 s (s.d.). This time range already indicates that under normal physiological conditions, the activation of RBCs is not compatible with an active process contributing to the initiation of a blood clot, but once caught in the fibrin network, the RBCs may actively support clot formation. In addition, to test the necessity of the presence of $Ca^{2+}$ during LPA stimulation, the negative control experiments excluded any interplay between the infrared trapping laser and the adhesion process.

## Signaling components

Because LPA is a phospholipid derivative, we examined the extent to which it is directly involved in the adhesion process. Although the concentration used was clearly below the critical micelle concentration (70 μM - 1 mM) [26], which might induce detergent-like effects, LPA is likely to be incorporated into the membrane. From the initial intercellular adhesion after LPA stimulation (Fig. 2b), one may propose the following alternative explanations: (i) LPA is directly responsible for the adhesion, (ii) LPA and the $Ca^{2+}$ influx are both necessary to mediate adhesion, and (iii) LPA simply triggers the $Ca^{2+}$ influx, and the $Ca^{2+}$ signaling alone is sufficient to induce adhesion. Option (i) can be excluded immediately because it was tested as a control in the initial set of experiments (Fig. 2b). In order to discriminate between options (ii) and (iii), experiments where A23187, a $Ca^{2+}$ ionophore, was added to the RBCs were performed. As shown in Fig. 3, the increased intracellular $Ca^{2+}$ concentration is the dominant signal initiating the adhesion. The $Ca^{2+}$ entry under LPA stimulation is channel-mediated, although the molecular identity of the channel remains unclear [29]. Because LPA is not the major entity in the adhesion process, an alternative molecule or a combination of several entities downstream of the $Ca^{2+}$ signal must control the response. Proteins that are known to be activated in RBCs by an increase in intracellular $Ca^{2+}$ concentration are the Gardos channel, the lipid scramblase, the cysteine protease calpain [30, 31] and the $Ca^{2+}$ pump. Although all the proteins are $Ca^{2+}$-activated, their sensitivity to $Ca^{2+}$ differs. To determine in which order and under what conditions the above mentioned players activate, we refer to the $Ca^{2+}$ concentration with the half maximal effect ($EC_{50}$). It would therefore be desirable to quantitatively measure the $Ca^{2+}$ concentration in an individual RBC. Unfortunately, this is not possible due to the failure of ratiometric $Ca^{2+}$ sensors in RBCs [28]. Instead, we compare $EC_{50}$ values determined in different studies with the cellular responses we observed in the present study. The smallest $EC_{50}$ for $Ca^{2+}$ is obviously that of the $Ca^{2+}$ pump that keeps the resting $Ca^{2+}$ concentration in RBCs well below 100 nM [32]. With any increase in the intracellular $Ca^{2+}$ concentration, the $Ca^{2+}$ pump will activate. Although the $V_{max}$ of the $Ca^{2+}$ pump was determined in cell populations to be patient- and sample-dependent in a range of 8 to 20 mmol/($l_{cells}$·h) [32], the pump activity per cell varies tremendously [33]. This variation explains the broad time range observed between the start of LPA stimulation and the $Ca^{2+}$ increase. While the pump can counterbalance the LPA-induced $Ca^{2+}$ influx for a short time period, during the application of A23187, the amount of $Ca^{2+}$ entering the cell exceeds the $V_{max}$ capacity of the $Ca^{2+}$ pump for all cells. At a $Ca^{2+}$ concentration of 400 nM, the flippase transporting PS actively from the outer membrane leaflet to the inner one is inhibited [34]. Once the $Ca^{2+}$ influx exceeds the transport capacity of the $Ca^{2+}$ pump, the first player that will be activated is the Gardos channel, with an $EC_{50}$ of 4.7 µM [35]. As we can see for the shape transitions upon A23187 stimulation shown in Fig. 3b, cells turn (transiently) into echinocytes as a consequence of KCl loss triggered by $K^+$ efflux through the Gardos channel. For LPA stimulation, the situation is different: due to the activation of the non-selective cation channel by LPA, a $Na^+$ influx and, consequently, NaCl uptake counterbalances the KCl loss initiated by the Gardos channel, producing an osmotic equilibrium. For a 2.5 µM LPA stimulation, this equilibrium [36] is reached, whereas for a 10 µM LPA stimulation, the effect of the NaCl uptake overwhelms the KCl loss, resulting in the formation of spherocytes. The next entity to be activated upon $Ca^{2+}$ entry is the scramblase, with an $EC_{50}$ of 29 µM [37]. Scramblase activity was demonstrated by probing for PS in the outer membrane leaflet using annexinV-FITC staining. Staining was present after both LPA and A23187 stimulation, as depicted in Figs. 3c and 3d, respectively. Calpain is activated with an $EC_{50}$ of 40 µM, which is very close to the $EC_{50}$ of the scramblase [38]. Under both stimulation conditions (LPA and A23187), we observed vesiculation that has been shown to be associated with the activation of calpain [30], which cleaves spectrin and actin and therefore leads to the breakdown of the cytoskeleton. This is in good agreement with a recent report on exovesiculation by Cueff et al. [39]. However, the vesiculation was much more pronounced under A23187 stimulation than under LPA stimulation, as depicted in the representative images in Fig. 3d and Fig. 2c, respectively. Therefore, we suggest that the $Ca^{2+}$ concentration after stimulation with 2.5 µM LPA is smaller compared to the A23187 stimulation and might be in the range of $EC_{50}$ of calpain. For A23187 stimulation, a shape change from echinocytes to spherocytes occurs (compare Figs. 3a and 3b), which is mediated by the encapsulation of microvesicles. However, the occurrence of PS on the outside of the cell makes it a good candidate for initiating the adhesion process. This could be due to simple $Ca^{2+}$-$PS^-$ cross-bridging and/or a more complex process involving adhesion proteins. Further evidence for both options comes from aggregation studies of PS vesicles [40, 41], where aggregation occurs in solutions of physiological ionic strength containing $Ca^{2+}$ in the mM concentration range. The dependence on high $Ca^{2+}$ concentrations and evidence from further studies reporting enhanced aggregation of PS liposomes in the presence of polymers [42] suggest that additional membrane constituents in the RBC contribute to the aggregation process. This leads to other $Ca^{2+}$-dependent proteins in RBCs, such as $PKC_\alpha$ [43, 44] or the nitric oxide synthase [45, 46]. Further research is required to address the question of the molecular identity of the additional components in the adhesion process. Analyzing the stepwise unbinding (compare Fig. 4b) when RBCs are separated from each other (corresponding to step 3 in Fig. 4a), we found a Gaussian distribution of the force centered at 71.9 pN (Fig. 4e). Such a distribution suggests the formation of tethers and specific bonds between the RBCs that are released one by one during the separation process.

Relevance to in vivo conditions

The LPA concentration of between 2.5 and 10 µM is a physiologically relevant concentration that is likely to occur locally after platelet activation. Upon stimulation with such an LPA concentration, RBCs adhere irreversibly to each other. The separation force of approximately 100 pN (determined by single cell force spectroscopy) is in a range that is of relevance in the vasculature [47]. As mentioned previously, due to the time course of the $Ca^{2+}$ increase, we regard an initiation of a blood clot based on intercellular RBC adhesion to be irrelevant under physiological conditions. However, once caught in the fibrin network of a blood clot, the adhesion process observed here in vitro may support the solidification of the clot. This notion is supported by the aforementioned experimental and clinical investigations reporting a prolongation of bleeding time in subjects with low RBC counts [7-9, 48]. Evidence that the adhesion process described in this paper may play a role in vivo was recently provided by Chung and coworkers [49]. In this study, an increase in intracellular $Ca^{2+}$ of RBCs associated with a PS exposure was be related to prothrombotic activity in vivo in a venous thrombosis rat model. Under pathophysiological conditions, intercellular RBC adhesion after $Ca^{2+}$ influx seems to have a more pronounced effect. An example is the vasco-occlusive crisis of sickle cell disease (SCD) patients. Here, the $Ca^{2+}$ influx is mediated by the NMDA-receptor, which has been found to be abundant in RBCs [46]. Our study provides a link between the increased prevalence of the NMDA-receptor in SCD patients [50] and the symptoms of the vasco-occlusive crisis. Further examples where disorders in the ion homeostasis of RBCs are associated with thrombotic events are malaria [51] and thalassemia [52, 53]. Therefore, we propose that the $Ca^{2+}$ increase, independent of the entry pathway, followed by PS exposure and RBC aggregation is a general mechanism that may become relevant under pathological conditions.

Acknowledgements

This work has been supported by the DFG Graduate School, GRK 1276 and the Ministry of Economy and Sciences of the Saarland. The study was approved by the ethics committee of the Medical Association of the Saarland (reference number 132/08).

## Conflict of Interest Disclosure

All authors have no conflicts of interest to declare.

Figures

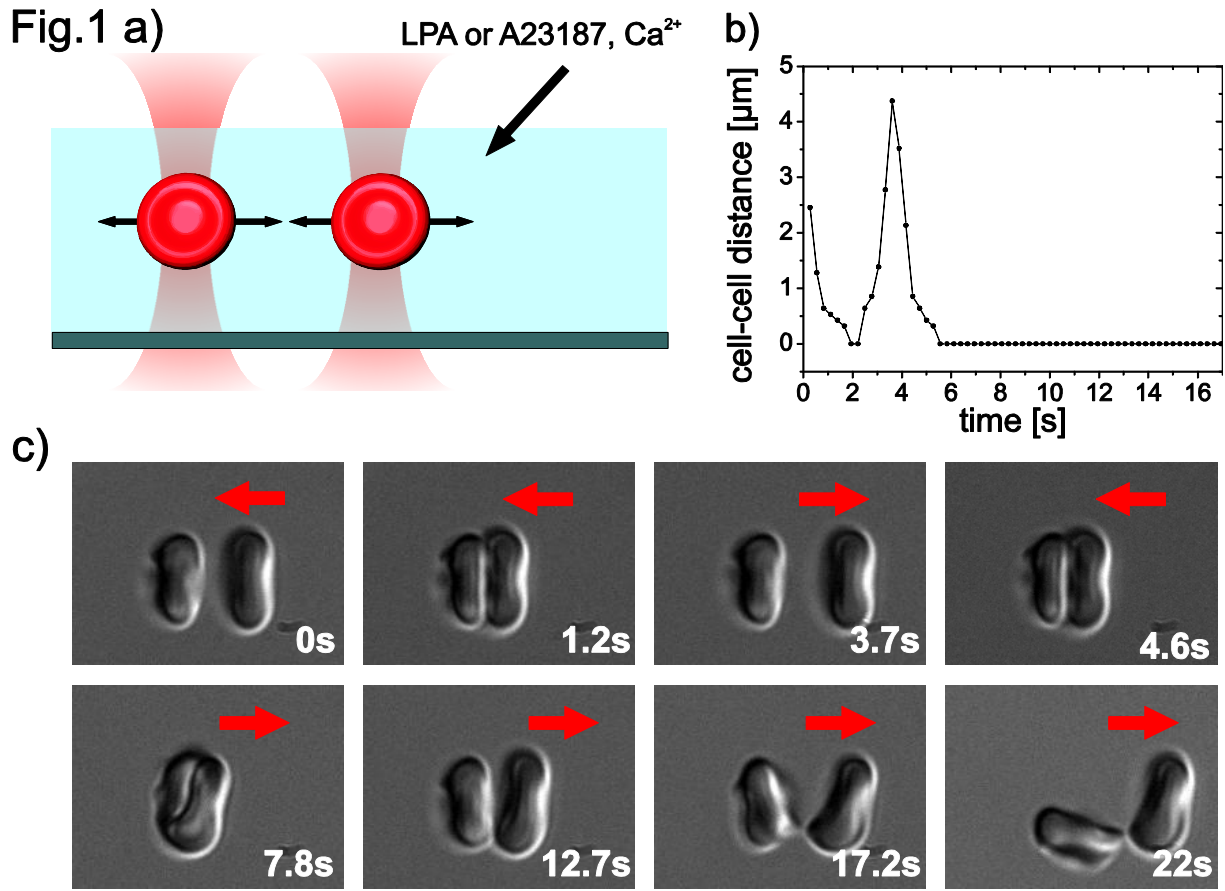

**Figure 1:** Probing for adhesion between RBCs after LPA stimulation. Panel a) represents a sketch of the oscillatory movement of two trapped cells. The actual cell-cell distance over the course of the adhesion test is depicted in panel b). The graph depicts the separation distance of two red blood cells, as determined using edge detection, over the course of the experiment. Cells were preincubated with 2.5 µM LPA for 5 min. After adhesion occurred at around 6 s, the distance between the cells remained zero. Panel c) depicts representative images of an adhesion measurement of two RBCs held by 4 optical traps. During a recording period of 26 s, the cells were moved back and forth as indicated by the red arrows. The full video can be seen in the supplemental material.

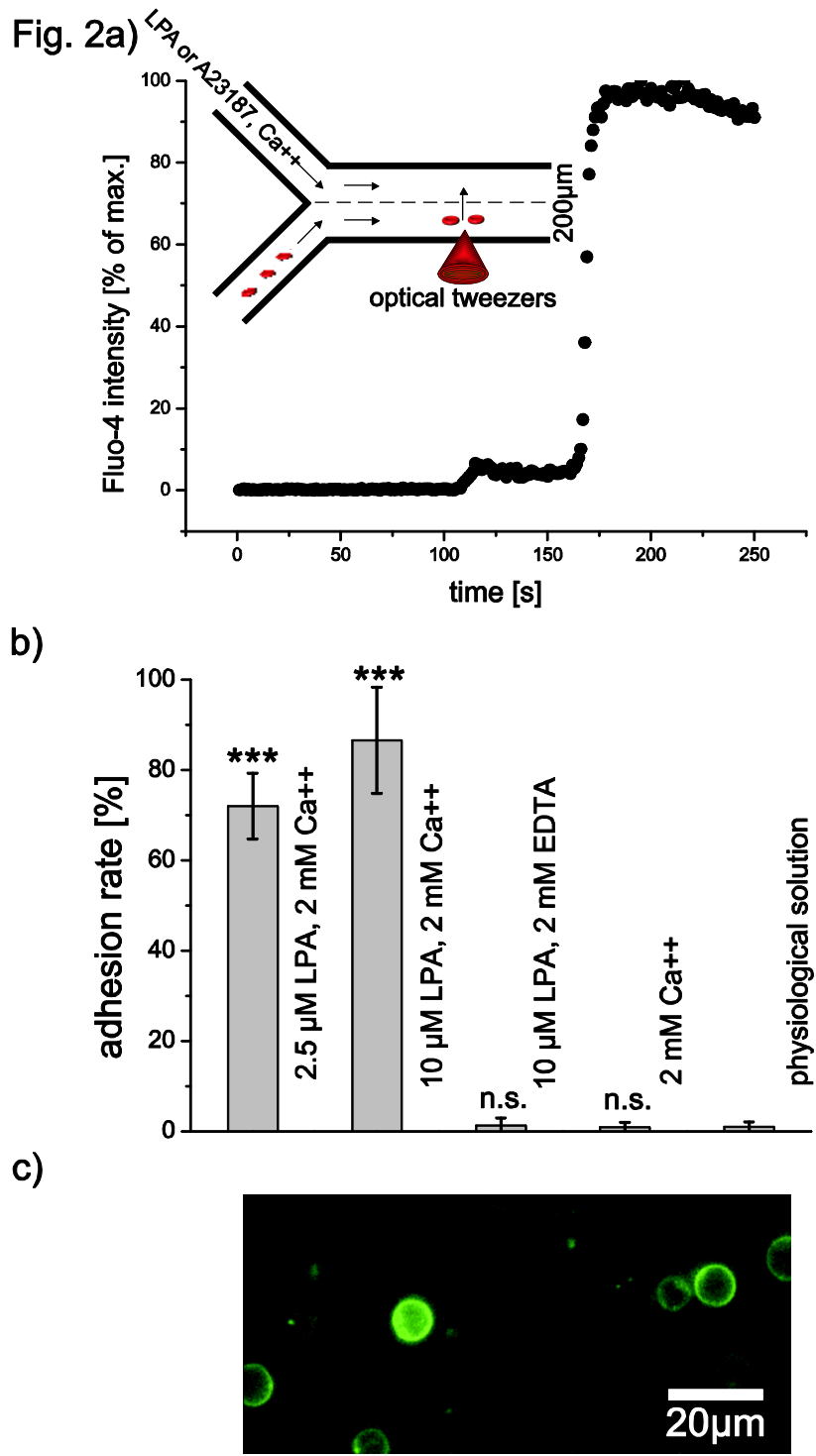

**Figure 2:** The response of RBCs to LPA. a) The relative fluorescence signal of a representative RBC in the upper microfluidic channel; t = 0 s is the time when the cell was moved into the LPA solution. The inset shows a schematic picture of the microfluidic chamber used. b) Results of the LPA measurements conducted in the microfluidic chamber and the Petri dish. The grey bars represent the percentage of cells that showed adhesion. The overall number of cells tested was 60 cells per measurement. In the presence of LPA and $Ca^{2+}$, a significant number of cells showed adhesion, whereas in the control experiments, only a very small portion of the cells showed an adhesion. The results of the Student's t-test, compared to the control measurement (PIS-solution), are indicated at the top of each bar. c) A fluorescence image of LPA-treated (2.5 µM) RBCs after annexin V-FITC

staining. The annexin V binding indicates the presence of PS on the outer membrane leaflet of the cells.

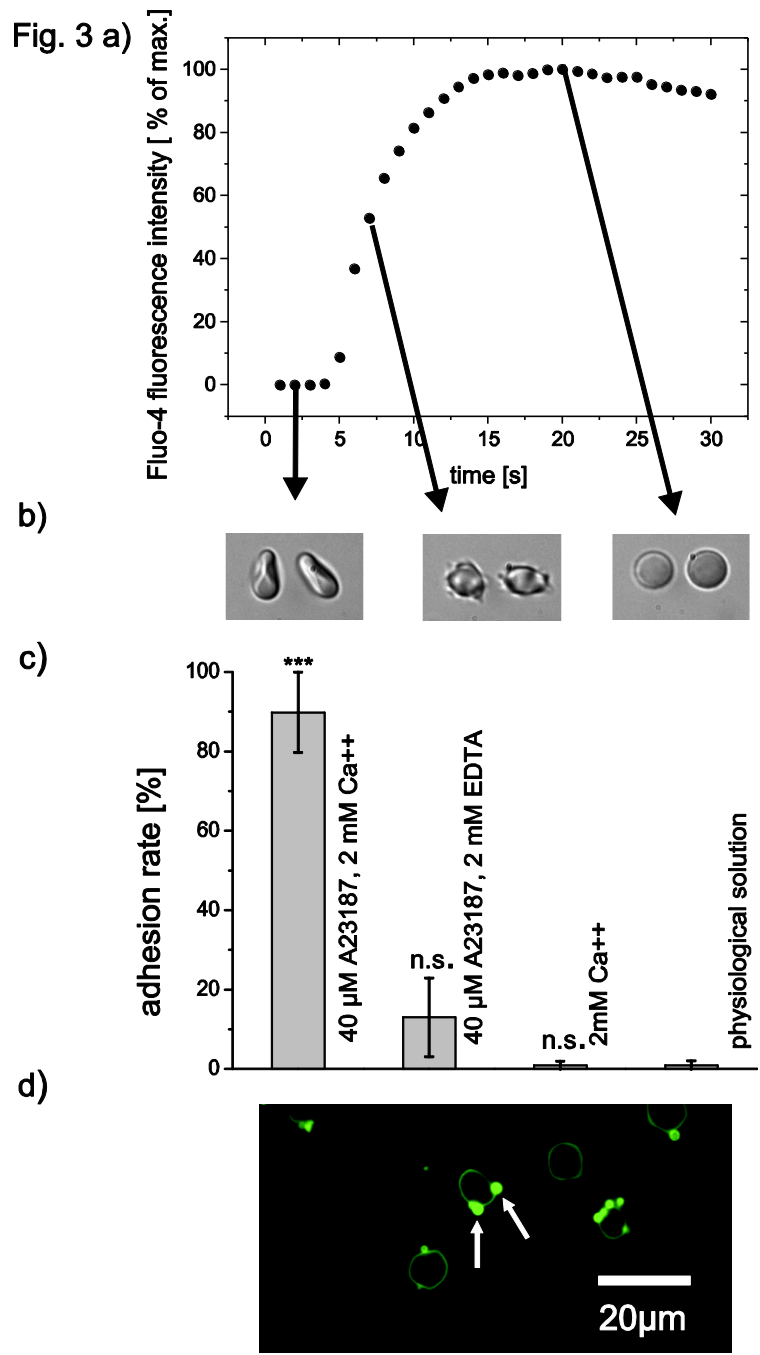

**Figure 3:** Measurements with the $Ca^{2+}$ ionophore A23187. a) The relative fluorescence signal of a representative RBC in the upper channel of the microfluidic system; t = 0 s is the time when the cell reached the A23187 solution. The $Ca^{2+}$ increase happens almost instantaneously. The decrease in signal after 15 s is due to photobleaching. b) The RBCs undergo a shape transformation from discocytes (left) via echinocytes (middle) to spherocytes (right) after transfer into the A23187 buffer solution in the upper channel. c) The results of the ionophore measurements conducted in the microfluidic chamber and a Petri dish. The black bars represent the percentage of cells that showed adhesion. The number of cells tested was about 60 per measurement. In the presence of A23187 and $Ca^{2+}$, about 90% of the cells tested adhered, whereas in the control experiments, less than 3% of the cells adhered. The results of the Student's t-test, compared to the control measurement (PIS-solution), are indicated at the top of each bar. d) A fluorescence image of annexin V-FITC-labeled RBCs. The cells were treated with A23187, and exposure of PS at the cell surface was clearly identified. A vesiculation of the cells was also observed (indicated by arrows).

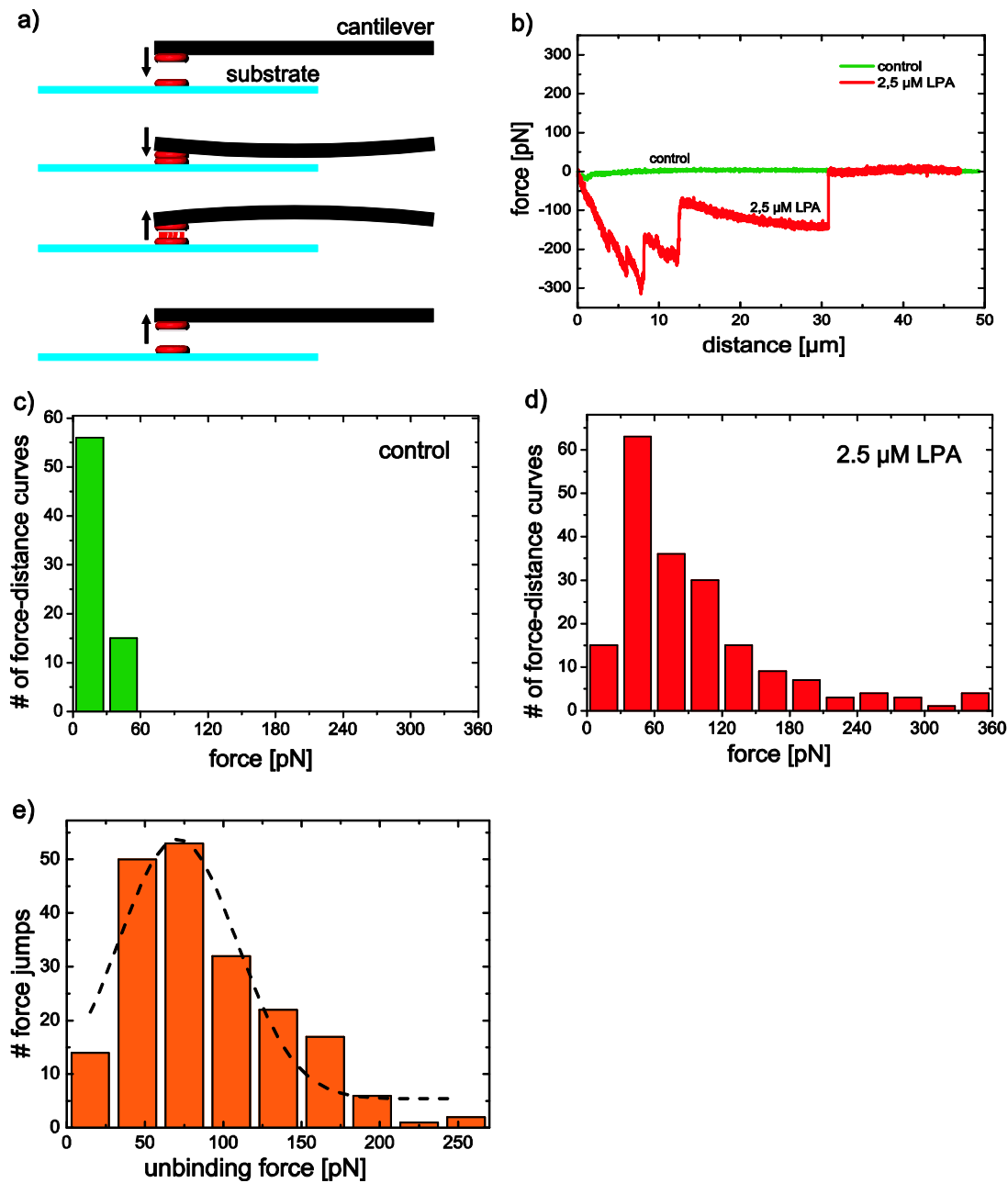

**Figure 4:** Force quantification using atomic force spectroscopy (AFS). Panel a) shows the working principle of the AFS. One cell is attached to a cantilever, while another one is attached to the surface. Over the course of the experiment, the cells are brought into contact and are withdrawn again. The adhesion force of the two cells can be measured by measuring the deflection of the cantilever. Panel b) shows the combined plot of an example force-distance-curve of a control (green) and an LPA measurement (red). Panel c) and panel d) depict the statistics of the measured forces in the controls and LPA measurements, respectively. Panel e) provides a histogram of the unbinding force of a single tether within the procedure visualized in panels a) and b). The dotted line depicts a Gaussian approximation of the bars.